# Comprehensive Electrical Control of Metamagnetic Transition of a Quasi-2D Antiferromagnet by in-situ Anisotropic Strain

*Han Zhang, Lin Hao, Junyi Yang, Josh Mutch, Zhaoyu Liu, Qing Huang, Kyle Noordhoek, Andrew F. May, Jiun-Haw Chu, Jong-Woo Kim, Philip J. Ryan, Haidong Zhou and Jian Liu*

Dr. H. Zhang, Dr. L. Hao, J. Yang, Q. Huang, K. Noordhoek, Prof. H. Zhou, Prof. J. Liu
Department of Physics and Astronomy, The University of Tennessee, 217A A.H. Nielsen Physics Bldg, Knoxville, TN 37996, USA
Email: jianliu@utk.edu

J. Mutch, Dr. Z. Liu, Prof. J.-H. Chu
Department of Physics, University of Washington, Seattle, WA, USA, 98195

Dr. A. F. May
Materials Science and Technology Division, Oak Ridge National Laboratory, Oak Ridge, TN, USA, 37831

Dr. J.-W. Kim, Dr. P. J. Ryan,
Advanced Photon Source, Argonne National Laboratory, Argonne, IL, USA, 60439

Dr. P. J. Ryan
School of Physical Sciences, Dublin City University, Dublin 11, Ireland

**Abstract:** Effective nonmagnetic control of the spin structure is at the forefront of the study for functional quantum materials. This study demonstrates that, by applying anisotropic strain up to only 0.05%, the metamagnetic transition field of spin-orbit-coupled Mott insulator $Sr_2IrO_4$ can be in-situ modulated by almost 300%. Simultaneous measurements of resonant x-ray scattering and transport reveal that this drastic response originates from the complete strain-tuning of the transition between the spin-flop and spin-flip limits, and is always accompanied by large elasto- and magneto-conductance. This enables electrically controllable and electronically detectable metamagnetic switching, despite the antiferromagnetic insulating state. The obtained strain-magnetic field phase diagram reveals that $C_4$-symmetry-breaking anisotropy is introduced by strain via pseudospin-lattice coupling, directly demonstrating the pseudo-Jahn-Teller effect of spin-orbit-coupled complex oxides. The extracted coupling strength is much weaker than the superexchange interactions, yet crucial for the spontaneous symmetry-breaking, affording the remarkably efficient strain-control.

Quantum materials with strong charge–spin-orbital-lattice interplay are known to exhibit exotic emergent phenomena, such as colossal magnetoresistance, unconventional superconductivity, and nontrivial topological phases [1]. Tuning one degree of freedom by another could lead to novel functional controls, which is particularly important for the emerging field of antiferromagnetic (AF) spintronics [2], where effective detection and control of the magnetic structure remains an outstanding question. Varying magnetic field and temperature are the two most common methods but not ideal for applications [3]. Fundamentally, the magnetic structure of AF order critically depends on spin anisotropy, the energy scale of which is however largely determined by spin-orbital-coupling (SOC) strength and lattice symmetry and is difficult to be externally controlled. Ferroelastic strain is recently reported to be able to switch the uniaxial magnetic anisotropy in the intermetallic $Mn_2Au$ [4]. Achieving such an efficient control in correlated quantum materials may lead to tuning not only the antiferromagnetism but also other competing orders.

$Sr_2IrO_4$ is one of the most important representatives of the spin-orbit-coupled correlated oxide. Its strong SOC stabilizes an AF Mott insulating ground state with the so-called $J_{eff}=1/2$ moments, which represents the effective angular momentum or pseudospin of the localized electrons [5]. Yet this pseudospin-half Mott state can be measured electronically as a semiconductor and exhibits large conductive responses [6] to modulations of the AF structure, giving rise to an excellent prototype of AF spintronics [5,7-9]. This state is highly unusual because the profound impacts of strong SOC are often hidden in the low-energy physics. For instance, due to the strong SOC in $Sr_2IrO_4$, one may anticipate significant magnetocrystalline anisotropy, which however turns out to be relatively weak within the *ab*-plane [10]. The large anisotropic exchanges caused by SOC are actually square-lattice symmetry-invariant and have no contribution to the magnetic anisotropy [11]. Indeed, $Sr_2IrO_4$ exhibits incredible phenomenological analogy to the weakly spin-orbit-coupled high-$T_c$ cuprates [12]. Due to the intrinsic isotropy of the $J_{eff}=1/2$ moments and their exchange interactions, the spontaneous rotational symmetry breaking of the magnetic phase transition in such a system must come from higher-order interactions, which are much weaker but vital for the magnetic anisotropy. In particular, model Hamiltonian calculations recently proposed that a so-called 'pseudo-Jahn-Teller effect' [13, 14], which is an extension of the famous Jahn-Teller effect of orbital degenerate systems to orbital singlet with strong entanglement with spin, is critical for spin-orbit-coupled oxides, such as $Sr_2IrO_4$. It predicts that the AF order necessarily causes lattice orthorhombicity, which in turn stabilizes a pseudospin easy axis. Such an emergent magnetoelastic coupling is expected to affect many properties, but a direct observation is lacking due to the small spontaneous orthorhombicity at $10^{-4}$.

Here we directly demonstrate the 'pseudo-Jahn-Teller' effect as a remarkably efficient inverse magnetoelastic coupling in controlling the AF order and the metamagnetic transition in $Sr_2IrO_4$. By a simultaneous measurement of x-ray resonant magnetic scattering (XRMS) and magneto-transport, we unambiguously show that an anisotropic strain less than 0.1%, applied onto the square lattice through a piezo device, is sufficient to in-situ tune the AF pseudospin axis and the metamagnetic transition among the spin-flop, spin-flip, and continuous regimes with large modulations in the elasto- and magneto-conductance. As a result, the metamagnetic transition field is shifted by almost 300%, enabling metamagnetic switching by electrical control and electronic detection. The in situ tunability allows us to establish the strain-magnetic field phase diagram of $Sr_2IrO_4$ for the first time, and reveal the $C_4$-symmetry-breaking anisotropy driven by the in situ strain as a quantitative measure of the pseudospin-lattice coupling strength, which is indeed small compared with the AF exchange interaction and yet highly decisive for the functional controls of correlated electrons with pseudospin.

**Figure 1a** shows the tetragonal crystal structure of $Sr_2IrO_4$ and its magnetic structures below $T_N$~230K [7]. While the $J_{eff}$=1/2 moments are antiferromagnetically ordered within each Ir atomic layer, the combination of SOC and octahedral rotation introduces large anisotropic exchanges and canting of the in-plane AF moments. The canted moments of different Ir planes further cancel each other out by stacking in a ↑-↑-↓-↓ sequence at zero field. This emergent collinear AF order of the canted moments of individual layers transitions into a ferromagnetic (FM) state when the external magnetic field overcomes the effective AF inter-layer coupling, leading to a metamagnetic transition accompanied with a giant magnetoconductance (MC) [6, 15], indicating a strong charge response that is unexpected in the picture of a strong Mott insulator. One may consider this as a spin-valve-like [16] behavior at the atomic scale, where the neighboring Ir planes with parallel canted moments pair up as one slab (↑-↑ ⇀ ⇑), and the parallel (⇑⇑) and antiparallel (⇑⇓) configurations of the adjacent slabs create two conductive values as the "on" and "off" states. For simplicity, the two configurations are respectively denoted by FM and AF states hereafter.

To introduce a symmetry-breaking anisotropic strain $\varepsilon$ within the Ir plane, we mounted the sample onto a piezo actuator [17] with maximum deliverable strain ~0.1%. Surprisingly, despite such small strain strength, we found a robust and almost complete on-and-off switch of the conductance (Figure 1c) when alternating the piezo voltage to stretch the sample along $a$-axis at 210 K with $H//b$ at 600 Oe. We further measured the actual strain induced in the sample with synchrotron x-ray diffraction (see methods) and found the maximum $\varepsilon$ ~0.05%. This elastoconductance (EC) response indicates a highly efficient in-situ strain-control of the metamagnetic transition by gently breaking the tetragonality of the layered structure. To shed light on this strain effect, we performed a systematic MC study (Figure 1d) at various strain values under the same geometry. A clear shift of the metamagnetic transition by hundreds of Oersteds can be immediately noticed. Figure 1e shows that the magnetic field $H_{FM}$, where the transition fully enters the FM state, decreases rapidly and systematically from 1500 to 600 Oe with $\varepsilon$ increasing to ~0.05%. That is a 250% increase as the strain decreases. One can see that the shape of the MR curve is also strongly distorted with strain. A complementary measurement by continuously varying $\varepsilon$ at fixed $H$ further reveals a highly nonlinear EC when crossing the transition (Figure S3-S4 in the supplementary).

The complex and drastic conductance responses indicate a nontrivial mechanism. In fact, since the strong charge response is unexpected for a typical Mott insulator, it is unclear whether and how the conductance changes can reflect the change of the underlying magnetic structure. To gain microscopic insights, we developed a sample environment for simultaneous XRMS and conductance measurements under in situ controls of both strain and magnetic fields (Figure 2a). Such an in situ multi-control-multi-detection technique allows unambiguous comparison between the conductance response and the metamagnetic transition in different strain and magnetic fields. In particular, the spontaneous $C_4$ rotational symmetry-breaking of the AF order necessarily leads to energetically equivalent domains where the AF axes are related to each other by 90º rotation. These twin domains may undergo different processes during the metamagnetic transition. At 210 K, we observed the magnetic peaks at the Ir $L_3$-edge corresponding to the two 90º AF twin domains at zero field with the canted moments along the *a*-axis ($AF_a$) and *b*-axis ($AF_b$), respectively. When applying *H//b* (Figure 2c), a detwinning process is first observed where the $AF_b$ domain gradually converts into the $AF_a$ domain. The magnetic peak of the FM state already emerges during this process, but increases significantly only when the $AF_a$ peak intensity begins to decrease until the completion of the metamagnetic transition. This magnetic detwinning process is consistent with a previous report [14]. Interestingly, when applying $\varepsilon$, both the $AF_b$-to-$AF_a$ and $AF_a$-to-FM processes are clearly sharpened, while being pushed toward higher and lower fields, respectively (Figures. 2d&e). The overall result is that the stable region of the $AF_a$ domain is strongly reduced and eventually diminished, driving the system into a new regime where the transition directly occurs between the $AF_b$ and FM state through a sharp jump (Figure 2f). In other words, the applied strain, albeit small, is able to detwin the AF domains as well but in a competing fashion against the magnetic detwinning, which in turn stabilizes the FM state at smaller $H$ and significantly shift $H_{FM}$, consistent with the MC data. Indeed, the simultaneously measured MC (Figure 2c-f) is found to always closely follow the modulation of the FM peak. The evolution from the two-step extended process of $AF_b$-to-$AF_a$-to-FM to the sharp single step of $AF_b$-to-FM can actually be seen in the MC curves when analyzing their field derivative (Figure 1d and Figure S3 in the supplementary). This correspondence reveals the nature of the shape change of the MC. The sensitivity of the conductance to the AF domain conversion is because the order parameter of the AF$_a$ domain already acquires a FM component evidenced by the finite FM peak intensity.

Given the conductance being a good measure of the transition processes, we constructed a $\varepsilon$-$H$ phase diagram and designated the AF domains and the FM state based on XRMS. The phase diagram in Figure 3a clearly illustrates that the broad transition region corresponding to the gradual AF$_a$-to-FM processes is suppressed by increasing $\varepsilon$ and eventually completely removed at $\varepsilon > 0.03\%$, giving rise to the downshifting and sharpening of the transition toward the AF$_b$-to-FM jump. These different magnetic structure evolutions represent two fundamentally different types of the metamagnetic transition, the spin-flip and spin-flop types [18]. The former involves a sudden flip of half of the AF spins when the magnetic field is parallel to the easy axis, whereas the latter undergoes spin rotation first to the hard axis and then gradually towards the field. The AF$_b$-to-FM jump in the phase diagram is essentially a spin-flip transition with the $b$-axis being the easy axis. On the other hand, the AF$_b$-to-AF$_a$-to-FM process corresponds to the spin-flop-type, where weak uniaxial anisotropy allows the slab magnetizations to rotate by 90º to the hard axis. Although the two slabs mostly remain antiparallel after this spin-flop step, it enables canting between the two slabs to accommodate the Zeeman energy gain in the expense of the anisotropy energy. The canting angle continuously increases until all the AF exchange energy is lost, evidenced by the observed gradual increase of the FM peak intensity at the expense of the AF$_a$ peak. It is remarkable that these two distinct transitions are reachable here within 0.05% of strain under in situ continuous tuning.

To quantitatively measure the energetics, we simulated the phase diagram with a minimum model of a two-slab cell, where the free energy can be written as:

$$F(\alpha_1, \alpha_2) = J_c \cos(\alpha_1 - \alpha_2) - h \cos(\alpha_1)$$
$$- h \cos(\alpha_2) - K_u(\cos^2\alpha_1 + \cos^2\alpha_2) - K_b [\cos(2\alpha_1) + \cos(2\alpha_2)]^2$$

where $J_c$ represents the inter-slab exchange energy; $\alpha_1$ and $\alpha_2$ are the angles of the slab magnetization *m* with respect to $H//b$-axis; $h = mH$ is the Zeeman energy. $K_u$ and $K_b$ represent the strain-induced uniaxial anisotropy and the bi-axial anisotropy due to $C_4$ symmetry, respectively. The simulated $K_u - h$ phase diagram in the unit of $J_c$ is shown in Figure 3b, and reproduces the overall experimental results very well with spin-flip and spin-flop phase boundaries, demonstrating the capability of the model in capturing the key physics. For instance, the relative angle between the two slabs is indeed continuously modulated within the AF$_a$ state. This model further predicts that a broad continuous transition would occur at large but negative $K_u$, which corresponds to an easy *a*-axis under a large negative *ε*. While negative *ε* is unreachable due to the voltage limit of the piezo actuator, we realize this scenario by simply applying $H\perp b$. We indeed observed a significant broadening of the metamagnetic transition into an extended gradual process with $H_{FM}$ increased to ~2300 Oe (Figure 3c), which is four times of the $H_{FM}$ ~ 600 Oe with $H//b$ under the same *ε*.

The ability of tuning the metamagnetic transition into all three regimes with a small *ε*<0.1% demonstrates the high efficiency and critical role of pseudospin-lattice interaction. One can see that $K_u$ is only ~10% of $J_c$ when entering the spin-flip regime, while $h/J_c$ is always ~1 near the transition. We estimated $J_c$~ 313 *n*eV at 210 K (supplementary), meaning that a $K_u$~31 $n$eV suffices to cover the full control of the transition. This enables quantification of the pseudospin-lattice interaction of $Sr_2IrO_4$ to be ~0.5 $m$eV/$\mu_B^2$ at 210 K, given that $\varepsilon$~0.05% and the staggered moment is locked to the canted moment. Since pseudospin-half moment is isotropic[10, 19], this anisotropic interaction couples the pseudospin quadrupole moment and the lattice as described in the pseudo-Jahn-Teller effect [13], and is much smaller than the Heisenberg superexchange interaction and the SOC-induced anisotropic exchange interactions [5, 20]. This is also consistent with the weak spontaneous orthorhombicity upon AF order. And yet this pseudo-Jahn-Teller mechanism is predominant in the symmetry-breaking of the ordering and provides a remarkably efficient control through the inverse effect, where the AF order responds to externally induced orthorhombicity. This response is robust in a wide temperature range as the anisotropic strain-induced shift of $H_{FM}$ remains similar when cooled down to 120 K (see Figure S5 in the supplementary).

To summarize, our results demonstrated that the application of in situ anisotropic strain is an appealing route to probe such rich spin-charge-lattice interplay of the spin-orbit-coupled 5$d$ electrons in static properties and even dynamic responses [21]. The development of in situ multifunctional sample environment is thus crucial. Indeed, our simultaneous measurement platform of XRMS and transportation property with in-situ strain and magnetic field controls at low temperatures is essential to establish the deterministic control of the AF domains and the metamagnetism as well as its close correlation with the complex electronic modulations. As such, our study is able to construct the strain-magnetic field phase diagram of $Sr_2IrO_4$ and directly demonstrate the pseudo-Jahn-Teller effect, opening a new avenue in the field of iridates and more broadly functional 5d transition metal oxides by playing emergent phenomena with *in situ* strain.

**Experimental Section**

See the Supporting Information for experimental details.

**Supporting Information**

Supporting Information is available from the Wiley Online Library or from the author.

**Acknowledgments**

J.L. and H.D.Z. acknowledge support from the Organized Research Unit Program at the University of Tennessee. Sample synthesis (A.F.M.) was supported by the U. S. Department of Energy, Office of Science, Basic Energy Sciences, Materials Sciences and Engineering Division. The in-situ strain control and measurement setup are partially supported by AFOSR DURIP Award FA9550-19-1-0180; the Scholarly Activity and Research Incentive Fund (SARIF) at the University of Tennessee and as part of Programmable Quantum Materials, an Energy Frontier Research Center funded by the U.S. Department of Energy (DOE), Office of Science, Basic Energy Sciences (BES), under award DE-SC0019443. Z.L. and J.H.C. acknowledge the support of the David and Lucile Packard Foundation.  Transport measurement is supported by the U. S. Department of Energy under grant No. DE-SC0020254 and the Electromagnetic Property (EMP) Lab Core Facility at the University of Tennessee. This research used resources of the Advanced Photon Source, a U.S. Department of Energy (DOE) Office of Science User Facility operated for the DOE Office of Science by Argonne National Laboratory under Contract No. DE-AC02-06CH11357. The authors thank David Mandrus, Mark P. M. Dean and Cristian Batista for valuable discussions; Randal R. McMillan and Bennett S. Waddell for providing technical support in making strain devices and sample holders.

**References**

[1] a) R. Schaffer, E. Kin-Ho Lee, B. J. Yang, Y. B. Kim, Rep. Prog. Phys. **2016**, 79, 094504; b) J. G. Rau, E. K.-H. Lee, H.-Y. Kee, Annu. Rev. Condens. Matter. Phys. **2016**, 7, 195; c) W. Witczak-Krempa, G. Chen, Y. B. Kim, L. Balents, Annu. Rev. Condens. Matter Phys. **2014**, 5, 57.

[2] a) T. Jungwirth, X. Marti, P. Wadley, J. Wunderlich, Nat. Nanotechnol. **2016**, 11, 231; b) V. Baltz, A. Manchon, M. Tsoi, T. Moriyama, T. Ono, Y. Tserkovnyak, Rev. Mod. Phys. **2018**, 90, 015005; c) P. Němec, M. Fiebig, T. Kampfrath, A. V. Kimel, Nat. Phys. **2018**, 14, 229; d) L. Šmejkal, Y. Mokrousov, B. Yan, A. H. MacDonald, Nat. Phys. **2018**, 14, 242.

[3] C. Song, Y. You, X. Chen, X. Zhou, Y. Wang, F. Pan, Nanotechnology **2018**, 29, 112001.

[4] X. Chen, X. Zhou, R.Cheng, C. Song, J. Zhang, Y. Wu, Y. Ba, H. Li., Y. Sun, Y. You, Y. Zhao and F. Pan, Nat. Matt. **2019,**18, 931.

[5] G. Jackeli, G. Khaliullin, Phys. Rev. Lett. **2009**, 102, 017205.

[6] a) M. Ge, T. F. Qi, O. B. Korneta, D. E. De Long, P. Schlottmann, W. P. Crummett, G. Cao, Phys. Rev. B **2011**, 84, 100402; b) C. Wang, H. Seinige, G. Cao, J. S. Zhou, J. B. Goodenough, M. Tsoi, Phys. Rev. X **2014**, 4, 041034; c) N. Lee, E. Ko, H. Y. Choi, Y. J. Hong, M. Nauman, W. Kang, H. J. Choi, Y. J. Choi, Y. Jo, Adv. Mater. **2018**, 30, e1805564.

[7] a) G. Cao, J. Bolivar, S. McCall, J. E. Crow, Phys. Rev. B **1998**, 57, R11039. b) B. J. Kim, H. Ohsumi, T. Komesu, S. Sakai, T. Morita, H. Takagi, T. Arima, Science **2009**, 323, 1329.

[8] G. Cao, P. Schlottmann, Rep. Prog. Phys. **2018**, 81, 042502.

[9] J. Bertinshaw, Y. K. Kim, G. Khaliullin, B. J. Kim, Annu. Rev. Condens. Matter. Phys. **2019**, 10, 315.

[10]    a) J. G. Vale, S. Boseggia, H. C. Walker, R. Springell, Z. Feng, E. C. Hunter, R. S. Perry, D. Prabhakaran, A. T. Boothroyd, S. P. Collins, H. M. Rønnow, D. F. McMorrow, Phys. Rev. B **2015**, 92; b) L. Fruchter, D. Colson, V. Brouet, J. Phys. Condens. Matter. **2016**, 28, 126003.

[11]    a) L. Hao, D. Meyers, H. Suwa, J. Yang, C. Frederick, T. R. Dasa, G. Fabbris, L. Horak, D. Kriegner, Y. Choi, J.-W. Kim, D. Haskel, P. J. Ryan, H. Xu, C. D. Batista, M. P. M. Dean, J. Liu, Nat. Phys. **2018**, 14, 806; b) S. Calder, D. M. Pajerowski, M. B. Stone, A. F. May, Phys. Rev. B **2018**, 98, 220402.

[12]    a) H. Watanabe, T. Shirakawa, S. Yunoki, Phys. Rev. Lett. **2010**, 105, 216410; b) F. Wang, T. Senthil, Phys. Rev. Lett. **2011**, 106, 136402; c) Y. K. Kim, N. H. Sung, J. D. Denlinger, B. J. Kim, Nat. Phys. **2015**, 12, 37; d) Y. J. Yan, M. Q. Ren, H. C. Xu, B. P. Xie, R. Tao, H. Y. Choi, N. Lee, Y. J. Choi, T. Zhang, D. L. Feng, Phys. Rev. X **2015**, 5, 041018; e) R. Comin, R. Sutarto, F. He, E. H. da Silva Neto, L. Chauviere, A. Frano, R. Liang, W. N. Hardy, D. A. Bonn, Y. Yoshida, H. Eisaki, A. J. Achkar, D. G. Hawthorn, B. Keimer, G. A. Sawatzky, A. Damascelli, Nat. Mater. **2015**, 14, 796.

[13]    H. Liu, G. Khaliullin, Phys. Rev. Lett. **2019**, 122, 057203.

[14]    J. Porras, J. Bertinshaw, H. Liu, G. Khaliullin, N. H. Sung, J. W. Kim, S. Francoual, P. Steffens, G. Deng, M. M. Sala, A. Efimenko, A. Said, D. Casa, X. Huang, T. Gog, J. Kim, B. Keimer, B. J. Kim, Phys. Rev. B **2019**, 99, 085125.

[15]    a) I. Fina, X. Marti, D. Yi, J. Liu, J. H. Chu, C. Rayan-Serrao, S. Suresha, A. B. Shick, J. Zelezny, T. Jungwirth, J. Fontcuberta, R. Ramesh, Nat. Commun. **2014**, 5, 4671; b) C. Wang, H. Seinige, G. Cao, J. S. Zhou, J. B. Goodenough, M. Tsoi, J. Appl. Phys. **2015**, 117; c) H. Wang, C. Lu, J. Chen, Y. Liu, S. L. Yuan, S.-W. Cheong, S. Dong, J.-M. Liu, Nat. Commun. **2019**, 10, 2280.

[16]    B. G. Park, J. Wunderlich, X. Marti, V. Holy, Y. Kurosaki, M. Yamada, H. Yamamoto, A. Nishide, J. Hayakawa, H. Takahashi, A. B. Shick, T. Jungwirth, Nat. Mater. **2011**, 10, 347.

[17]    J.-H. Chu, J. Hsueh-Hui Kuo, G. Analytis, I. R. Fisher, Science **2012**, 337, 710.

[18]    E. Stryjewski, N. Giordano, Adv. Phys. **1977**, 26, 487.

[19]    a) S. Fujiyama, H. Ohsumi, T. Komesu, J. Matsuno, B. J. Kim, M. Takata, T. Arima, H. Takagi, Phys. Rev. Lett. **2012**, 108, 247212; b) S. Calder, J. W. Kim, A. E. Taylor, M. H. Upton, D. Casa, G. Cao, D. Mandrus, M. D. Lumsden, A. D. Christianson, Phys. Rev. B **2016**, 94, 220407.

[20]    J. Kim, D. Casa, M. H. Upton, T. Gog, Y. J. Kim, J. F. Mitchell, M. van Veenendaal, M. Daghofer, J. van den Brink, G. Khaliullin, B. J. Kim, Phys. Rev. Lett. **2012**, 108, 177003.

[21]    a) S. Bahr, A. Alfonsov, G. Jackeli, G. Khaliullin, A. Matsumoto, T. Takayama, H. Takagi, B. Büchner, V. Kataev, Phys. Rev. B **2014**, 89, 180401; b) Y. Gim, A. Sethi, Q. Zhao, J. F. Mitchell, G. Cao, S. L. Cooper, Phys. Rev. B **2016**, 93, 024405.

**Figure 1**

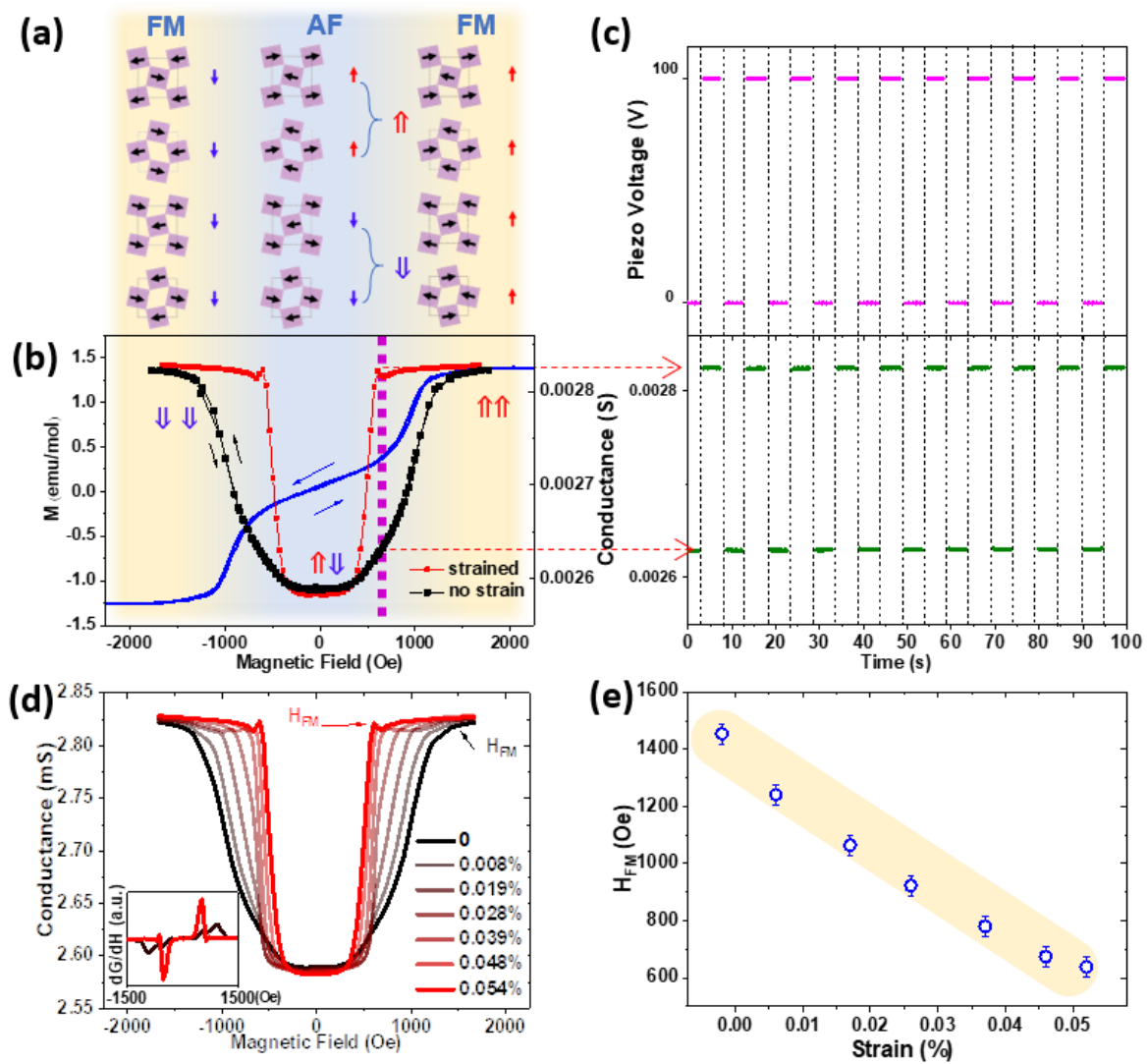

**Figure 1** (a) Pseudospin configuration of the AF state (middle) and the corresponding FM structure under positive (right) and negative (left) magnetic field along the *b*-axis. The pseudospins are canted by ~12°, owing to the strong SOC and octahedral rotation. It should be noted that, strictly speaking, all the states are canted AF orders of the pseudospins. (b) Magnetoconductance (MC) curve of $Sr_2IrO_4$ (black) with no anisotropic strain (black) and strained by a piezo actuator under 100 V (red). The blue curve is the in-plane magnetization measured at 210 K without strain, the metamagnetic transition takes place at ~1000 Oe, where a significant increase is seen in both the in-plane magnetization and the MC curve. For instance, a positive MC of ~10% is observed within ±250 Oe of the transition region. (c) Switching of the MC at 600 Oe by tuning the in-situ anisotropic strain along the *a*-axis (i.e., $\varepsilon//a$) while fixing $H//b$ near the transition. (d) Magnetoconductance under various anisotropic strains. (inset) 1$^{st}$ Derivative of MC under 0.008% and 0.054% anisotropic strain. (e) Shift of the field $H_{FM}$ as a function of strain where the metamagnetic transition is completed.

**Figure 2**

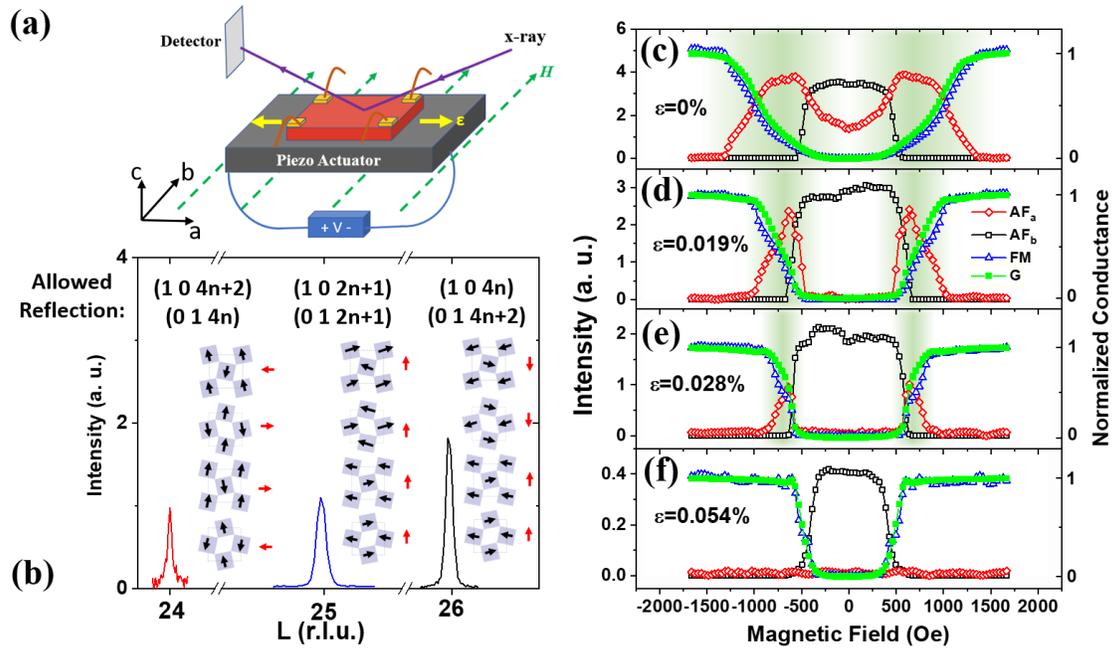

**Figure 2** (a) Schematic setup of the measurement. (b) The $AF_a$ (left), $AF_b$ (right) and FM (middle) structures and the corresponding magnetic reflections at 210K. (c-f) Magnetic field dependent magnetic reflection intensities at 210K with strain ranging from 0% to 0.054% along the *a*-axis. Red diamonds are the (0 1 24) reflection associated with $AF_a$, black squares are the (0 1 22) reflection from $AF_b$, the blue triangles are the FM peak (0 1 25), and the green curve is the in-situ conductance measurement.

**Figure 3**

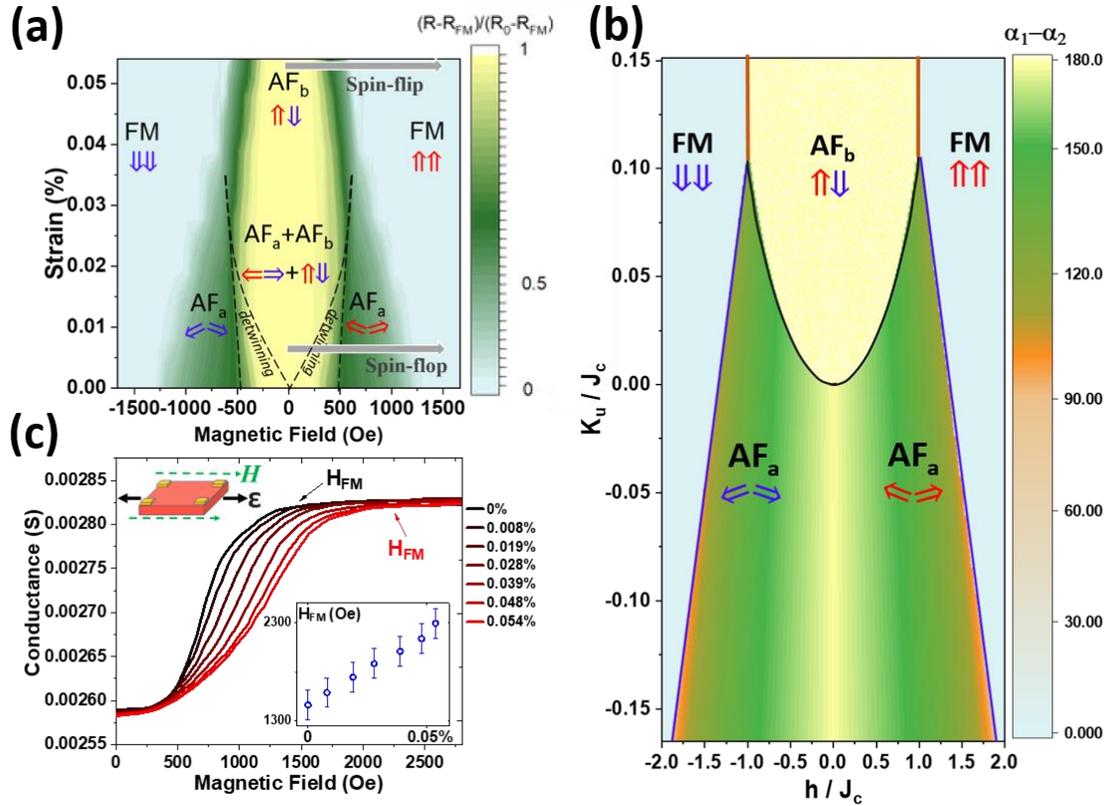

**Figure 3** (a) Experimental and (b) calculated (with $K_b=0.06J_c$) phase diagram of the $Sr_2IrO_4$ at 210K with varying anisotropic strain and magnetic fields. With negative strain value $H//b$ would be along the hard axis and the metamagnetic transition would be dramatically broadened. The simulated result in the spin flip regime (straight line) is a robust behavior regardless the value used for $K_b$. (c) Measured MC curves (solid lines) with both magnetic field and anisotropic strain along the *a*-axis. The inset shows the extract point $H_{FM}$ where the metamagnetic transition is completed with different strain values.